\documentclass[aps,prd,twocolumn,fleqn,superscriptaddress,preprintnumbers,showpacs]{revtex4-1}

\usepackage{graphicx,color}
\usepackage{amsmath,amssymb,amsfonts}
\usepackage{hyperref}

\newcommand{\be}{\begin{equation}}
\newcommand{\ee}{\end{equation}}

\newcommand{\bea}{\begin{eqnarray}}
\newcommand{\eea}{\end{eqnarray}}

\newcommand{\half}{{1\over2}}
\newcommand{\msbarbig}{{\overline{\mbox{\rm{MS}}}}}
\newcommand{\msbar}{{\overline{\mbox{\rmi{MS}}}}}

\newcommand{\ba}{\begin{eqnarray}}
\newcommand{\ea}{\end{eqnarray}}

\newcommand{\lpa}{\left(} 
\newcommand{\rpa}{\right)} 

\def\be{\begin{equation}}
\def\ee{\end{equation}}
\def\beq{\begin{eqnarray}}
\def\eeq{\end{eqnarray}}

\def\half{{\textstyle \frac 12}}

\def\gE{g_3}

\def\openone{\rlap 1\kern 0.22ex 1}
\newcommand{\rmi}[1]{{\mbox{\scriptsize #1}}}

\begin{document}

\input amssym.def
\input amssym.tex

\title{Quark number susceptibilities from resummed perturbation theory}

\preprint{BI-TP 2012/41}

\author{Jens O.~Andersen}
\affiliation{Department of Physics, Norwegian University of Science and Technology, N-7491 Trondheim, Norway}
\author{Sylvain Mogliacci}
\affiliation{Faculty of Physics, University of Bielefeld, D-33615 Bielefeld, Germany}
\author{Nan Su}
\affiliation{Faculty of Physics, University of Bielefeld, D-33615 Bielefeld, Germany}
\author{Aleksi Vuorinen}
\affiliation{Faculty of Physics, University of Bielefeld, D-33615 Bielefeld, Germany}

\begin{abstract}
We evaluate the second and fourth order quark number susceptibilities in hot QCD using two variations of resummed perturbation theory. On one hand, we carry out a one-loop calculation within hard-thermal-loop perturbation theory, and on the other hand perform a resummation of the four-loop finite density equation of state derived using a dimensionally reduced effective theory. Our results are subsequently compared with recent high precision lattice data, and their agreement thoroughly analyzed.
\end{abstract}

\pacs{12.38.Cy, 12.38.Mh}

\maketitle

\section{Introduction}
One of the most pressing challenges in the equilibrium thermodynamics of QCD is to develop nonperturbative tools to access the region of nonzero quark densities, addressing questions such as the existence and location of a critical point in the phase diagram. Barring a solution to the sign problem of lattice QCD, the leading method to determine the finite density equation of state (EoS), i.e.~the behavior of the pressure as a function of quark chemical potentials $\mu$, is through the evaluation of quark number susceptibilities,
\ba
\chi_{ijk...}(T)&\equiv& \frac{\partial^n p(T,\{\mu_f\})}
{\partial\mu_i\, \partial\mu_j \, \partial\mu_k\cdots}\bigg|_{\mu_f=0}\, ,
\label{suscdef}
\ea
where the indices $i,j,k,...$ refer to different quark flavors. These functions carry information about the response of the system to nonzero density, yet can be determined on the lattice without problems; for examples of recent studies, see e.g.~\cite{HotQCD,Wuppertal} and references therein. The applicability of these results to the determination of the EoS at $\mu\neq 0$ is ultimately restricted only by the convergence of the expansion of the pressure in powers of $\mu/T$.

While a quantitative description of the quark gluon plasma near its transition temperature $T_c$ clearly necessitates the use of nonperturbative techniques, it is also interesting to study, to what extent the behavior of the quark number susceptibilities can be understood using analytic weak coupling methods. First, unlike lattice calculations, perturbation theory works optimally at very high temperatures and thus offers a way to connect the results obtained around $T_c$ to arbitrarily high energies. More importantly, perturbative calculations are easily generalizable to finite density, and are not constrained to the region of small $\mu/T$. And finally, due to the cancelation of the \textit{purely} gluonic contributions to quark number susceptibilities, these quantities are expected to display improved convergence properties in comparison with the pressure itself.

Indeed, extensive analytic work on susceptibilities, and more generally the chemical potential dependence of the pressure, has been carried out within 
unresummed perturbation theory~\cite{av1,av2}, the hard-thermal-loop (HTL) approximation~\cite{Blaizot,Blaizot2,Chakraborty:2001kx,Chakraborty:2003uw,Haque:2011iz}, the analytically tractable large-$N_f$ limit of QCD~\cite{Ipp1,Ipp2}, and even the gauge/gravity duality \cite{Jorge}. In addition to this, the applicability of dimensional reduction (DR) to finite densities has been investigated in~\cite{Ipp3}, and the behavior of the susceptibilities determined through a nonperturbative DR study in~\cite{Hietanen}.

While many of the perturbative calculations listed above showed reasonably good agreement with lattice results existing at the time of their publication, the numerically significant corrections present in recent high precision lattice data~\cite{HotQCD,Wuppertal} clearly call for a re-examination of the issue. On top of this, the past years have witnessed important progress in the resummation of high-temperature perturbation theory on multiple fronts. In HTL perturbation theory (HTLpt)~\cite{lo-ym}, a recent evaluation of the partition function of hot QCD up to three-loop order has demonstrated dramatically improved convergence properties~\cite{nnlo-ym,nnlo-qcd}, and the agreement between the HTLpt and lattice results has subsequently been observed to be very good down to $2-3\,T_c$ (for the relevant lattice data, see e.g.~\cite{lat3,lat1,lat2}). In addition, the same framework has been applied to the case of finite density and zero temperature, albeit at lower orders~\cite{baier-redlich,hdlpt}.  At the same time, it was shown in~\cite{BIR,Laine} that a simple resummation of the soft, three-dimensional contributions to the four-loop EoS of hot QCD~\cite{Kajantie} is enough to considerably decrease its renormalization scale dependence, resulting in excellent agreement with lattice data. It should be interesting to see, what kind of an effect these new techniques have when applied to the evaluation of quark number susceptibilities.

In the present paper, our objective is simple: We want to apply state-of-the-art resummation techniques to the determination of the second and fourth order quark number susceptibilities, and compare the results to the most recent lattice data available. To this end, we address two separate calculations: First, we employ HTLpt to determine the susceptibilities at one-loop order, and after this apply the resummation scheme of \cite{BIR,Laine} to the four-loop finite density EoS of \cite{av1} to obtain ${\mathcal O}(g^6\ln\,g)$ results for the same quantities (dubbed `DR' in the following). Both calculations are performed within the $\msbarbig$ renormalization scheme, denoting the scale parameter by $\bar{\Lambda}$. When discussing the results, we will specialize to the phenomenologically most relevant case of three dynamical quark flavors, which for simplicity are all taken to be massless. We have, however, explicitly verified that keeping the leading order strange quark mass dependence in the results only 
affects them in any noticeable way at the very lowest temperatures.

\section{HTL perturbation theory}
Hard-thermal-loop perturbation theory is a reorganization of the usual perturbative expansion of thermal QCD. The Lagrangian density of the theory is written in the form
\ba
{\cal L_{\rm HTLpt}}&=&
\left(
{\cal L}_{\rm QCD}+{\cal L}_{\rm HTL}
\right)\Big|_{g\rightarrow\sqrt{\delta}g}
+\Delta{\cal L}_{\rm HTL}\,,
\ea
where ${\cal L}_{\rm QCD}$ is the undeformed Lagrangian of the theory, ${\cal L}_{\rm HTL}$ an HTL improvement term, and $\delta$ a formal expansion parameter introduced for bookkeeping purposes. The last part of the above expression, $\Delta{\cal L}_{\rm HTL}$, on the other hand contains counterterms, which are necessary to cancel the ultraviolet divergences introduced by the HTLpt reorganization.

For full QCD with dynamical quarks, the (gauge invariant) HTL improvement term reads
\ba\nonumber
{\cal L}_{\rm HTL}
&=&
-{1\over2}(1-\delta)m_D^2
{\rm Tr}\left(
F_{\mu\alpha}\bigg\langle{y^{\alpha}y^{\beta}\over(y\cdot D)^2}\bigg\rangle_{\!\!\!y}
F^{\mu}_{\beta}
\right)\\
&&+ (1-\delta)i\sum_f^{N_f} m^2_{q,f}\bar{\psi}_f\gamma^{\mu}
\bigg\langle{y_{\mu}\over y\cdot D}\bigg\rangle_{\!\!\!y}\psi_f\,,
\ea
where $D^{\mu}=\partial^{\mu}-igA^{\mu}$ denotes a covariant derivative (in the appropriate representation), $y=(1,\hat{\bf y})$ is a light-like four-vector, $\langle\rangle_y$ represents an average over the direction of $\hat{\bf y}$, and $m_D$ and $m_{q,f}$ are the Debye mass and fermion thermal mass parameters. Note that $m_{q,f}$ carries a dependence on the flavor index $f$, running from 1 to $N_f=3$.

HTLpt is formally defined as an expansion of physical quantities in powers of $\delta$ around $\delta=0$, implying that already at its leading order one is dealing with dressed propagators that incorporate important plasma effects, such as Debye screening and Landau damping. The starting point of HTLpt is thus an ideal gas of massive quasiparticles, which can be identified as one of the main reasons for its success. At higher orders, the expansion in $\delta$ generates dressed vertices as well as higher order terms that ensure that there is no overcounting of Feynman diagrams. 

In practice, physical observables are calculated within HTLpt by truncating the $\delta$-expansions at some specified order, and then setting $\delta=1$. If it were possible to carry out the expansion to all orders, the final result would be independent of the parameters $m_D$ and $m_{q,f}$. At any finite order in $\delta$, some residual dependence on them however remains, and a prescription for choosing their values is required. Optimally, the parameters should be determined via a variational condition for the thermodynamic potential, which is however only well defined beyond the leading order due to the absence of the coupling constant in the LO thermodynamic potential~\cite{nnlo-qcd-long}. In our calculation, we therefore identify the Debye and fermion masses with their weak coupling values,
\bea
m_D^2 &=& {g^2 \over 3} \bigg[\Big(N_c + {N_f \over 2}\Big) \, T^2 + \frac{3}{2\pi^2}\sum_f \mu_f^2 \bigg] \,, \\
m_{q,f}^2 &=& {g^2 \over 4} \frac{N_c^2-1}{4N_c} \bigg( T^2+\frac{\mu_f^2}{\pi^2}\bigg) \, ,
\eea
where we have kept the number of colors $N_c$ arbitrary. 

After the definitions above, the one-loop HTLpt determination of the EoS follows to a large extent the $\mu=0$ calculation of \cite{lo-ym}, including an analytic expansion of the result in powers of $m_{D}/T$ and $m_{q,f} /T$ up to order $g^5$. This results in
\begin{eqnarray}
p_\rmi{HTLpt}  &=& \frac{d_A\pi^2 T^4}{45} 
\Bigg\{ 1+\frac{N_c}{d_A}\sum_f\lpa\frac{7}{4}+30\hat{\mu}_f^2 +60\hat{\mu}_f^4\rpa \nonumber \\
& & -\frac{15}{2}\hat{m}_D^2 -  \frac{30N_c}{d_A}\sum_f\lpa 1+12\hat{\mu}_f^2\rpa  \hat{m}_{q,f}^2\nonumber \\
& &  +30\hat{m}_D^3 +\frac{60N_c}{d_A} \lpa 6-\pi^2\rpa \sum_f \hat{m}_{q,f}^4 \nonumber \\
& & +\frac{45}{4}\lpa \gamma_{\mbox{\tiny $E$}}-\frac{7}{2}+\frac{\pi^2}{3}+\log\frac{\bar{\Lambda}}{4\pi T}\rpa\hat{m}_D^4 \nonumber \\
& & + {\cal O}(\hat{m}_D^6,\hat{m}_{q,f}^6)\Bigg\} \, ,
\end{eqnarray}
where we have denoted $d_A\equiv N_c^2-1$ as well as introduced the dimensionless parameters $\hat{m}\equiv \frac{m}{2\pi T}$, etc. Results for the quark number susceptibilities are finally obtained by taking derivatives of this expression with respect to the chemical potentials, and setting $\mu=0$ in the end.

\section{Dimensional reduction}
To date, the unresummed weak coupling expansion of the QCD pressure has been determined up to and partially including its four-loop order, both at zero density~\cite{Kajantie,3dpert,3dnonpert} and at $\mu\neq 0$~\cite{av2}. A useful tool in these calculations has turned out to be the three-dimensional effective theory electrostatic QCD (EQCD), the partition function of which very conveniently encompasses the contributions of the soft and ultrasoft momentum scales ($gT$ and $g^2T$, respectively) to the corresponding quantity in the full theory \cite{Braaten:1995cm,Kajantie:1997tt}. In practice, one writes the pressure of the four-dimensional theory in the form
\ba
p_\rmi{QCD}&=& p_\rmi{HARD} + T\, p_\rmi{EQCD}\, ,
\ea
where $p_\rmi{HARD}$ is defined as the \textit{strict loop expansion} of the pressure in the full theory, obtained by letting dimensional regularization regulate both the UV and IR divergences. At the same time, $p_\text{EQCD}$ corresponds to the partition function of EQCD, which one can evaluate using a combination of perturbative \cite{3dpert,3dnonpert} and nonperturbative \cite{3dnonpert2,3dnonpert3} tools.

Formally, EQCD is a three-dimensional SU($N_c$) Yang-Mills theory coupled to an adjoint Higgs field $A_0$, originating from the zero Matsubara mode of the four-dimensional temporal gauge field. The theory is defined by the Lagrangian density
\begin{eqnarray}
    { \mathcal L}_{\rmi{EQCD}}
    &=&  
    \gE^{-2}
    \bigg\{
    \half \,{\rm Tr}[F_{ij}]^2
    + {\rm Tr}\!\big[(D_iA_0)^2\big]
    + m_{\rm E}^2 \, {\rm Tr}[A_0^2] \nonumber \\
    &&+i\zeta\, {\rm Tr}[A_0^3] + \lambda_{\rm E} \, {\rm Tr} [A_0^4]
    \bigg\}
    + \delta{\mathcal L}_\rmi{E} \,,
\label{LEQCD}
\end{eqnarray}
where we have assumed $N_c=3$ (for larger $N_c$ we would have two independent quartic terms for the $A_0$ field), and where the last term $\delta{\mathcal L}_\rmi{E}$ stands for a series of higher order non-renormalizable operators that start to contribute to the pressure only beyond ${\mathcal O}(g^6)$. The theory is parametrized by four constants: The three-dimensional gauge coupling $\gE$, the electric screening mass $m_{\rm E}$, the cubic coupling $\zeta \sim \sum_f \mu_f$ (see \cite{Hart:2000ha} for details), as well as the quartic coupling $\lambda_{\rm E}$. All of these parameters have expansions in powers of the four-dimensional gauge coupling $g$, and their values have been determined to the accuracy required by the four-loop evaluation of the EoS, some even beyond~(see e.g.~\cite{York}).

As discussed in~\cite{Laine}, the above way of writing the QCD pressure suggests a highly natural resummation scheme: While the unresummed weak coupling expansion is obtained by expanding the (perturbatively determined) EQCD partition function in powers of the four-dimensional gauge coupling $g$, one may simply skip the last step. This amounts to keeping $p_\rmi{EQCD}$ an unexpanded function of the effective theory parameters and writing
\ba
T\, p_\rmi{EQCD}&=& p_\rmi{M}+p_\rmi{G},
\ea 
where the functions $p_\rmi{M}$ and $p_\rmi{G}$ can be read off from eqs.~(3.9) and (3.12) of \cite{av2}. In~\cite{Laine}, this procedure was observed to lead to a considerable improvement in the convergence and scale dependence properties of the full theory pressure at zero chemical potential. It can, however, be applied to the case of the finite density pressure and the quark number susceptibilities with equal ease, which is what we have implemented in our calculations. An important step in this in principle straightforward exercise is to use the effective theory parameters in a form, where they have been analytically expanded in powers of $\mu/T$; cf.~Appendix D of \cite{av2} and Appendix B of \cite{Vuorinen:2004rd}. We refrain from writing the resulting, very long expressions here, but rather give them in the Mathematica file DREoS.nb  \cite{mathfile}. It is important to note that unlike in \cite{Laine}, in our calculation the unknown part of the ${\mathcal O}(g^6)$ term in the expansion of the pressure 
has not been fitted to lattice results, but simply been set to 0.

\section{Choice of parameters}
Before proceeding to a quantitative comparison of our results with lattice data, let us briefly discuss, how we have chosen the values of the parameters appearing in our calculations. These include the renormalization scale $\bar{\Lambda}$ as well as the QCD scale $\Lambda_{\msbar}$, in addition to which a prescription for determining the value (and running) of the gauge coupling must be specified. In all of these cases, we follow standard choices used widely in the literature.

In perturbative calculations of bulk thermodynamic observables, the renormalization scale $\bar{\Lambda}$ is typically first given a value of roughly $2\pi T$, around which it is then varied by a factor of 2 in order to test the sensitivity of the result with respect to the choice. Within DR, a commonly used prescription is to choose the central value by applying the Fastest Apparent Convergence (FAC) criterion to the three-dimensional gauge coupling $\gE$, resulting in $\bar{\Lambda}_\rmi{central}\approx 1.445\times 2\pi T$ \cite{Kajantie:1997tt}. For simplicity, we use this value in both of our computations.

For the dependence of the gauge coupling constant on the renormalization scale, we use a one-loop perturbative expression in the HTLpt result and a two-loop one in the DR case. This is in accordance with the usual rule that the uncertainties originating from the running of the gauge coupling should not exceed those due to the perturbative computation itself. Finally, for the choice of the QCD scale $\Lambda_\msbar$, we use a recent lattice determination of the strong coupling constant at a reference scale of 1.5\,GeV \cite{Bazavov:2012ka}. Requiring that our one- and two-loop couplings agree with this, we obtain the values $\Lambda_\msbar = 176$ and 283\,MeV in the two cases, respectively. To be conservative, we vary the value of the parameter around these numbers by 30\,MeV, which is somewhat larger than the reported lattice error bar.

\section{Results}

\begin{figure}[t]
\includegraphics[scale=0.34]{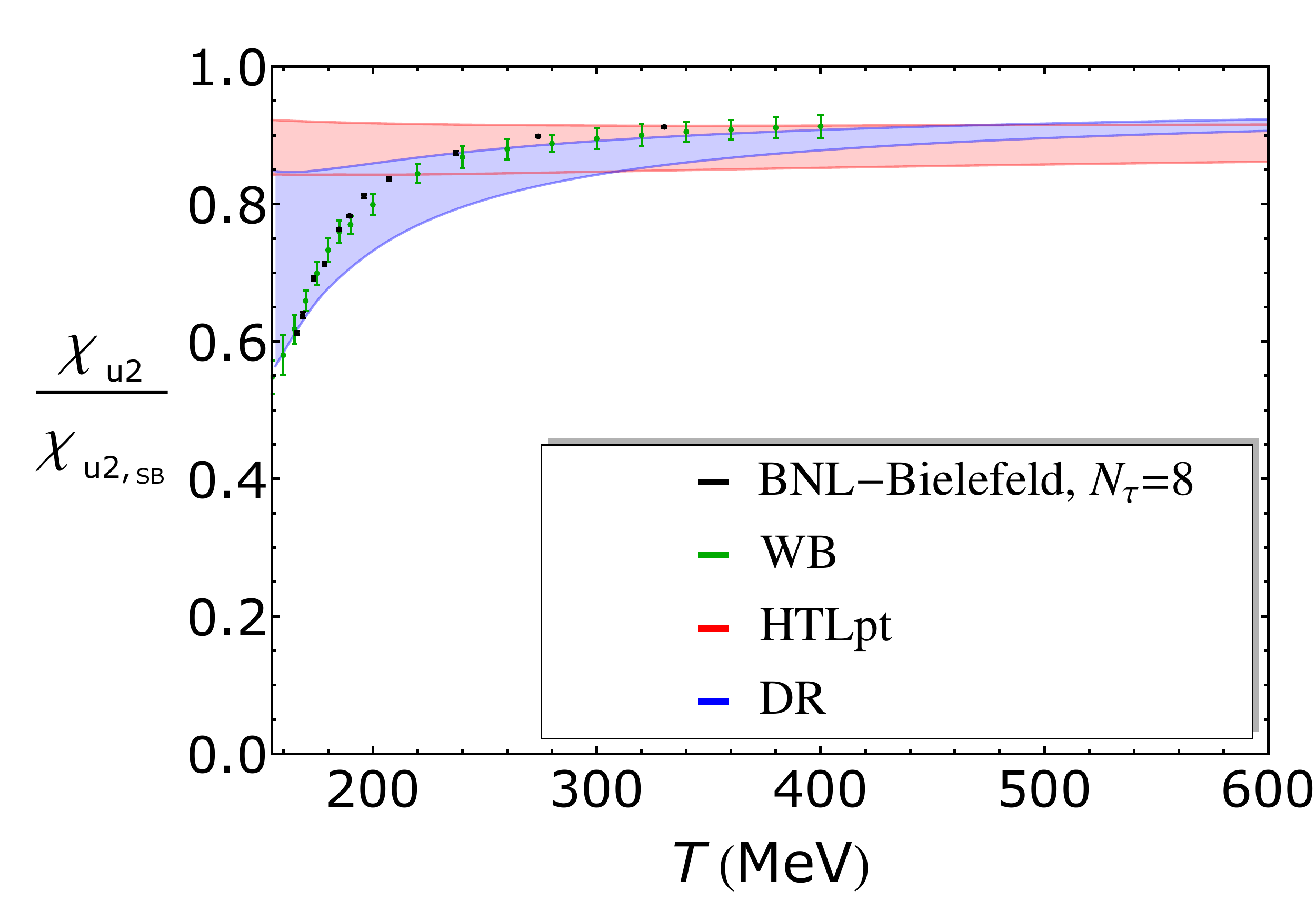}
\caption{A comparison of our HTLpt (red band) and DR (blue) results for the second order light quark number susceptibility $\chi_{\rm u2}$ with the recent lattice results of the BNL-Bielefeld~\cite{christian1,christian2} (black dots) and Wuppertal-Budapest (WB)~\cite{Wuppertal} (green) collaborations. All results have been normalized by the noninteracting Stefan-Boltzmann limit, while the bands corresponding to the perturbative results originate from varying the values of $\bar{\Lambda}$ and $\Lambda_\msbar$ within the ranges indicated in the text.
}
\label{nf3}
\end{figure}

In Fig.~\ref{nf3}, we finally display our predictions for the second order light quark number susceptibility $\chi_{uu}\equiv \chi_{\rm u2}$, normalized by the noninteracting Stefan-Boltzmann (SB) limit $\chi_{\rm u2, SB}=T^2$. The results are subsequently compared with the recent $N_\tau=8$ lattice data of the BNL-Bielefeld collaboration, obtained using the HISQ action~\cite{christian1,christian2}, as well as with the continuum extrapolated results of the Wuppertal-Budapest (WB) collaboration~\cite{Wuppertal}. As the widths of the red and blue bands --- corresponding respectively to the HTLpt and DR calculations --- demonstrate, the dependence of our results on the renormalization scale and the value of $\Lambda_\msbar$ is rather mild. For instance, a comparison of the DR band with the unresummed four-loop result of \cite{av1} shows a reduction of the uncertainty by a factor larger than 5 in this temperature range. Our two results are in addition in impressive agreement with each other at temperatures of the order of 300\,MeV and higher, deviating significantly only in the direct vicinity of $T_c$. The results are in addition seen to be in good agreement with lattice data, with the DR band even reproducing the decreasing trend of the lattice results at small $T$.

\begin{figure}[t]
\includegraphics[scale=0.34]{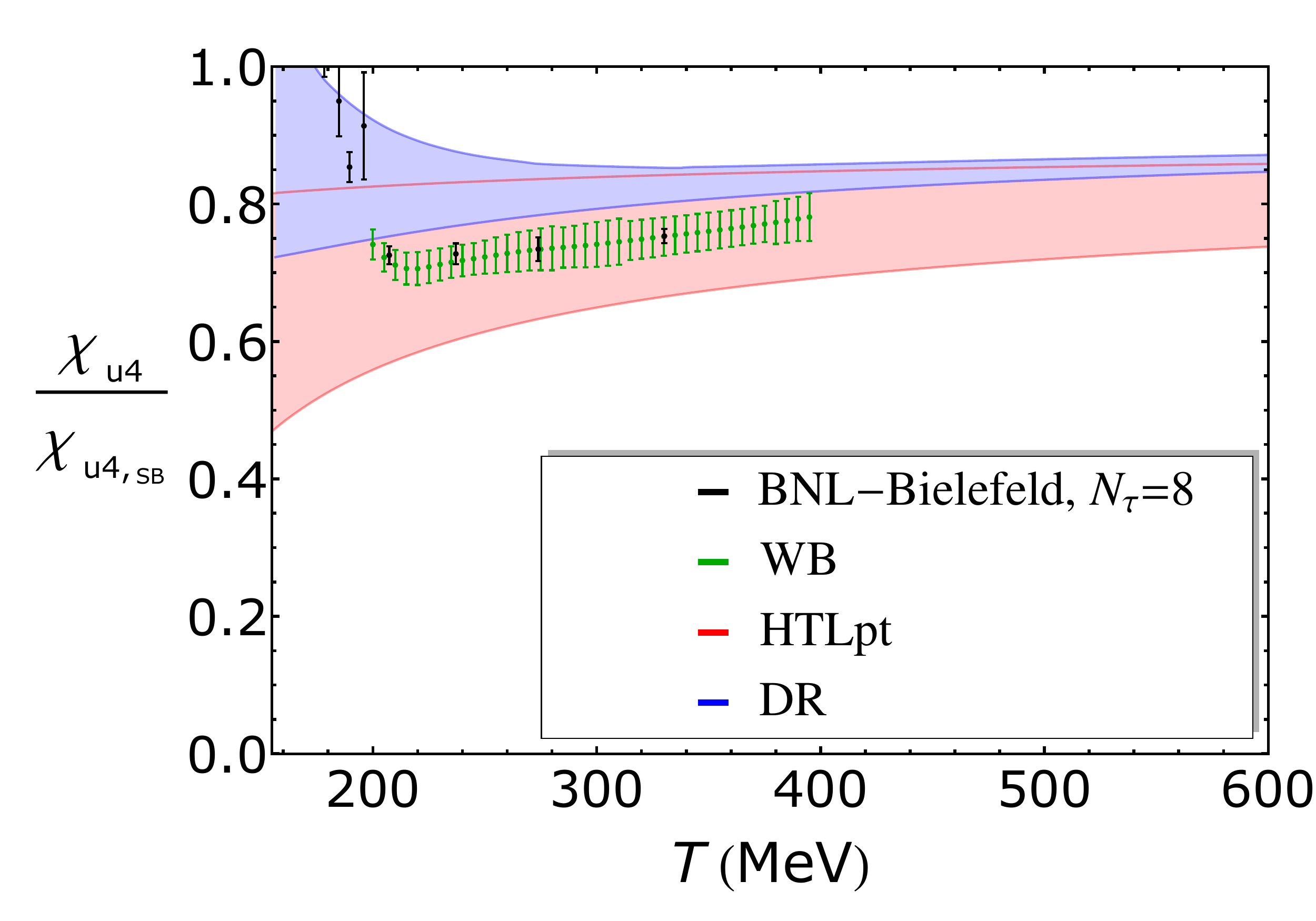}
\caption{As in Fig.~\ref{nf3}, but for the fourth order light quark number susceptibility $\chi_{\rm u4}$. The WB lattice data has been taken from \cite{szabolcs}.
}
\label{nf3b}
\end{figure}

Moving on to the fourth order susceptibilities, in Fig.~\ref{nf3b} we show our results for the quantity $ \chi_{uuuu}\equiv \chi_{\rm u4}$, also scaled by the corresponding SB value $\chi_{\rm u4, SB}=6/\pi^2$. The continuum extrapolated WB lattice data are this time taken from \cite{szabolcs}, while the $N_\tau=8$ BNL-Bielefeld results are again from~\cite{christian1,christian2}. Both data sets are seen to reside inside our HTLpt band down to $T\approx$ 200\,MeV, and in fact almost coincide with its central value on this interval. This fact may, however, be to some extent coincidental, considering its dependence on our (rather arbitrary) choice of $\bar{\Lambda}_\rmi{central}$. The DR prediction is again seen to reproduce the qualitative trend of the lattice results, but is observed to slightly overestimate them in the relevant temperature range. This disagreement, however, slowly decreases with temperature, and it is plausible that the DR band and the lattice error bars start to overlap already at 
temperatures below 500\,MeV once new data for higher temperatures emerges.

\begin{figure}[t]
\includegraphics[scale=0.34,trim=0 5 0 30, clip]{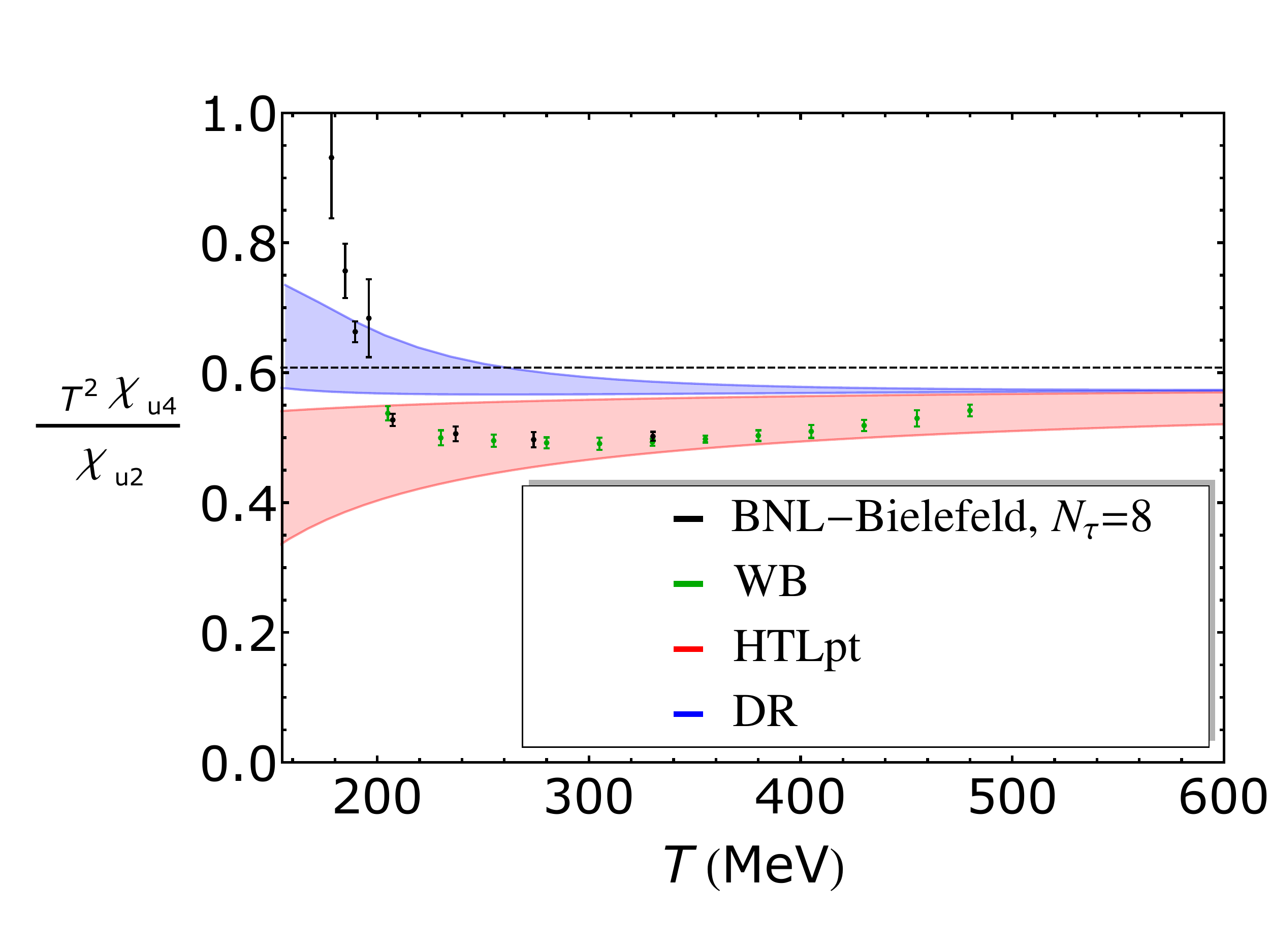}
\caption{The ratio of the results shown in the previous two figures. This time, the quantity is not normalized by the SB value, but rather approaches the value $6/\pi^2$ (dashed line) at high temperatures.}
\label{ratio}
\end{figure}

To highlight the difference between our two perturbative calculations, in fig.~\ref{ratio} we consider the ratio of the fourth and second order susceptibilities, for which much of the dependence of our results on the renormalization scale and $\Lambda_\msbar$ cancels. Indeed, the HTLpt and DR predictions for this quantity are seen to be highly robust, and in addition in disagreement with each other for all temperatures considered. The HTLpt result is observed to be consistent with the lattice data down to temperatures close to 200\,MeV, even though it does not reproduce the increasing trend of the latter close to $T_c$. At the same time, the DR band resides above the lattice data for most of the interesting temperature range, and while displaying a modest increase at low temperatures, is clearly not consistent with the lattice measurements. Considering the increase of the lattice data at temperatures close to 500\,MeV, it would nevertheless be of some interest to see whether they continue to favor the HTLpt prediction in the interval of 500-1000\,MeV; this issue remains to be decided by future lattice simulations.

The physical origins of the behavior described above are clearly interesting to analyze. Inspecting the DR result at different orders of the weak coupling expansion, it is seen to consistently improve both in the sense of approaching the lattice data and in exhibiting a decreasing dependence on $\bar\Lambda$ and $\Lambda_\msbar$. The fact that for the fourth order susceptibility the lattice points lie outside the displayed perturbative band for most temperatures may of course appear troublesome and indicative of an underestimation of the systematic uncertainties in the calculation; if so, this can clearly be attributed to the resummation performed, which has a dramatic effect on the size of the error bars. At the same time, the apparent success of our HTLpt result should be taken with some reservations, considering that it is only a leading order one. It has after all been repeatedly seen in perturbative calculations that the transition from LO to NLO may shift the result considerably and sometimes even 
increase its renormalization scale dependence. Indications of such behavior have indeed been recently reported in \cite{Haque:2012my,Haque:2013qta}.

\section{Conclusions}
In the present paper, we have applied two types of resummed perturbation theory to the determination of the second and fourth order light quark number susceptibilities in thermal QCD. Our main results are shown in Figs.~\ref{nf3}, \ref{nf3b}, and \ref{ratio}, of which in particular the last one, displaying the ratio of the two quantities, shows an interesting pattern. It is observed that the lattice data agree with our one-loop HTLpt result over a wide range of temperatures, while there is a slight, yet visible discrepancy between them and the four-loop DR result below $T\approx 500$\,MeV. It is obviously an important task to attempt to explain this observation, and in particular see, whether the present success of HTLpt still prevails once higher order corrections are included.

Clearly, the most important virtue of weak coupling methods is their versatility. Indeed, as soon as the  quark number susceptibilities for the three-flavor case treated here have been computed, it is straightforward to extend the results to other theories of interest, such as two-flavor or quenched QCD (or even QCD with a different number of colors), as well as to other quantities, such as higher order susceptibilities or the pressure as a function of chemical potentials. All of these cases, as well as a further study of the $N_f=3$ results displayed above, are examples of directions we will pursue in a forthcoming publication \cite{future}. Our hope is these results will eventually find phenomenological use in the study of the current and future heavy ion data from RHIC, LHC and FAIR.


\section*{Acknowledgments}
We are grateful to Szabolcs Borsanyi, Keijo Kajantie, Frithjof Karsch, Mikko Laine, Peter Petreczky, Anton Rebhan, Kari Rummukainen, Christian Schmidt, York Schr\"oder, Sayantan Sharma, Michael Strickland and Mathias Wagner for useful discussions. In addition, we would like to thank Szabolcs Borsanyi and Frithjof Karsch for providing us with their latest lattice data. S.~M.~and A.~V.~are supported by the Sofja Kovalevskaja program and N.~S.~by the Postdoctoral Research Fellowship of the Alexander von Humboldt Foundation.


\end{document}